\documentclass[12pt]{article}
\usepackage{amsmath,amsfonts,amssymb}
\begin{document}
\thispagestyle{empty}

\def\theequation{\arabic{section}.\arabic{equation}}
\def\a{\alpha}
\def\b{\beta}
\def\g{\gamma}
\def\d{\delta}
\def\dd{\rm d}
\def\e{\epsilon}
\def\ve{\varepsilon}
\def\z{\zeta}

\def\B{\mbox{\bf B}}
\newcommand{\eq}[1]{\begin{equation} #1 \end{equation}}
\newcommand{\al}[1]{\begin{align} #1 \end{align}}
\newcommand{\ml}[1]{\begin{multline} #1 \end{multline}}
\newcommand{\arr}[1]{\begin{array} #1 \end{array}}
\def\cp{\mathbb {CP}^3}
\def\o{\omega}
\def\p{\partial}
\def\rb{\right}
\def\lb{\left}
\def\sinh{\mathrm{sinh}}
\def\cosh{\mathrm{cosh}}
\def\arctanh{\mathrm{arctanh}}

\begin{titlepage}
\strut\hfill TUW--08--09

\renewcommand{\thefootnote}{\fnsymbol{footnote}}
\begin{center}
{\bf\Large Neumann-Rosochatius integrable system\\
for strings on $AdS_4\times\cp$}
\end{center}
\vskip 1.2cm \centerline{\bf Changrim  Ahn$^1$, P. Bozhilov$^1$
\footnote{On leave from Institute for Nuclear Research and Nuclear
Energy, Bulgarian Academy of Sciences, Bulgaria.} and R.C.
Rashkov$^2$ \footnote{On leave from Dept. of Physics, Sofia
University, Bulgaria.}} \vspace*{0.6cm} \centerline{\sl
$^1$Department of Physics} \centerline{\sl Ewha Womans University}
\centerline{\sl DaeHyun 11-1, Seoul 120-750, S. Korea}
\vspace*{0.6cm} \vspace*{0.3cm} \centerline{\sl $^2$ Institute for
Theoretical Physics} \centerline{\sl Vienna University of
Technology} \centerline{\sl Wiedner Haupsstr. 8-10, 1040 Vienna,
Austria} \vspace*{0.6cm} \centerline{\tt ahn@ewha.ac.kr,
bozhilov@inrne.bas.bg, rash@hep.itp.tuwien.ac.at}

\vskip 20mm

\baselineskip 18pt

\begin{center}
{\bf Abstract}
\end{center}
We use the reduction of the string dynamics on $AdS_4\times\cp$ to
the Neumann-Rosochatius integrable system.
All constraints can be expressed simply in terms of a few parameters.
We analyze the giant magnon and single spike solutions on $R_t\times\cp$
with two angular momenta in detail and find
the energy-charge relations. The finite-size effects of the giant magnon and single spike
solutions are analyzed.

\end{titlepage}
\newpage
\baselineskip 18pt

\def\nn{\nonumber}
\def\tr{{\rm tr}\,}
\def\p{\partial}
\newcommand{\bea}{\begin{eqnarray}}
\newcommand{\eea}{\end{eqnarray}}
\newcommand{\bde}{{\bf e}}
\renewcommand{\thefootnote}{\fnsymbol{footnote}}
\newcommand{\be}{\begin{equation}}
\newcommand{\ee}{\end{equation}}
\newcommand{\h}{\hspace{0.5cm}}

\vskip 0cm

\renewcommand{\thefootnote}{\arabic{footnote}}
\setcounter{footnote}{0}

\setcounter{equation}{0}
\section{Introduction}

The AdS/CFT correspondence \cite{Maldacena,GKP,Witten}, which has led
to many exciting developments in the duality between type IIB string
theory on $AdS_5\times S^5$ and ${\cal N}=4$ super Yang-Mills theory,
is now being extended into $AdS_4/CFT_3$.  A most
promising candidate is ${\cal N}=6$ super Chern-Simons theory
with $SU(N)\times SU(N)$ gauge symmetry and level $k$.  This model,
which was first proposed by Aharony, Bergman, Jafferis, and Maldacena
\cite{ABJM}, is believed to be dual to M-theory on $AdS_4\times
S^7/Z_k$.  Furthermore, in the planar limit of $N, k\to\infty$ with a
fixed value of 't Hooft coupling $\lambda=N/k$, the ${\cal N}=6$ Chern-Simons is
believed to be dual to type IIA superstring theory on $AdS_4\times\cp$.

Quantum integrability of the planar ${\cal N}=6$ Chern-Simons theory was first discovered
by Minahan and Zarembo in the leading two-loop-order
perturbative computation of the anomalous dimensions of
gauge-invariant composite operators \cite{MZ}. (See also \cite{BR}.)
Its excitation spectrum and symmetry have been studied in \cite{GGY}
and all-loop Bethe ansatz, first conjectured in \cite{GV}, was confirmed by
the exact $S$-matrix first proposed in \cite{AN}.

Integrability in the string theory side is also under active investigation.
The Penrose limit of the type IIA string and BMN-like spectrum have been
studied in \cite{NisTak}.
Various aspects of classical integrability in the $\lambda>>1$ limit have been
found in \cite{AruFro,Stef,FreGra,GroVieii}.
The giant magnon (GM) solution \cite{GGY,Grignani} and its finite-size effect
\cite{Grignanif,APGHO} have been computed.
The GM and single spike (SS) solutions of membranes on
$AdS_4\times S^7$ background and their finite-size effects have been
worked out in \cite{ABii}
and such string solutions as circular and pulsating strings \cite{CW} and
spiky strings and finite-size effects on $AdS_4\times\cp$ \cite{LPP} have been found.
Also recently, one-loop quantum correction to the GMs has been computed
\cite{Shend}.

All these solutions, however, are restricted to
the strings moving in $R_t\times S^2\times S^2$ with one angular momentum.
The configuration in the target space is in such a way that
the azimuthal angle of the string coordinates in the first $S^2$ is opposite to
that in the second sphere.
The purpose of this article is to find classical solutions with two angular momenta in
 $\cp$. Our string solutions develop spikes in the two spheres $S^2\times S^2\subset \cp$
with a certain dynamics in $U(1)$ fiber. The picture is  analogous to the
dyonic GM in $AdS_5\times S^5$ \cite{Dorey}.
We want to emphasize that the Neumann-Rosochatius (NR) integrable system is
very effective for dealing with the strings on $\cp$.
This integrable system is obtained by reformulating
the problem in a conformal gauge using the Polyakov action and
assuming a particular ansatz for string coordinates.
This approach has been previously developed and applied to
find classical solutions such as the GM \cite{HM} and SS \cite{SS}
of type IIB string theory on $AdS_5\times S^5$ in \cite{AFRT,ART,KRT06}.
The application of the NR system to the SS in $S^3$ has been worked out in
\cite{BobR07}, the most general case of GM and SS
has been considered in \cite{Dimov:2007ey} and to the finite-size effects in \cite{ABi}.
The NR integrable system in $AdS_4/CFT_3$ was used to find the GM solution
for the membrane on $AdS_4\times S^7$ \cite{MNR} and to compute the finite-size
effects in \cite{ABii}.
In this article we apply this system to the strings moving in the
$R_t\times\cp$ background. The space $\cp$ can be thought as a $U(1)$ fibration
over $S^2\times S^2$ (see the appendix for basic facts about $\cp$).
Our ansatz for string coordinates allows motion in
$S^2\times S^2$ subspace and $U(1)$ fiber as well.
The solutions we find contain the GM solutions for motion in $S^2\times S^2$ found in \cite{Grignani}
as a special case.
Using this formulation, we compute the finite-size effects of the GM and
the SS strings.

The paper is organized as follows. In section 2 we introduce the
classical string action on $R_t\times\cp$ and the corresponding
NR system. We provide explicit GM and SS solutions moving
in $R_t\times\cp$ and provide an analysis of the finite-size effects in section 3.
We conclude in Section 4 with a brief discussion of our results.

\setcounter{equation}{0}
\section{Strings on $R_t\times\cp$ and the NR Integrable System}

Let us start with the Polyakov string action \bea &&S^P=
-\frac{T}{2}\int d^2\xi\sqrt{-\gamma}\gamma^{mn}G_{mn},\h G_{mn} =
g_{MN}\p_m X^M\p_nX^N,\\ \nn &&\p_m=\p/\p\xi^m,\h m,n = (0,1),
\h(\xi^0,\xi^1)=(\tau,\sigma),\h M,N = (0,1,\ldots,9),\eea and
choose {\it conformal gauge} $\gamma^{mn}=\eta^{mn}=diag(-1,1)$, in
which the Lagrangian and the Virasoro constraints take the form
\bea\label{l}
&&\mathcal{L}_s=\frac{T}{2}\left(G_{00}-G_{11}\right) \\
\label{00} && G_{00}+G_{11}=0,\qquad G_{01}=0.\eea where
$T$ is the string tension.

The background metric $g_{MN}$ for $AdS_4\times\cp$ is given by
\bea\nn ds^2=g_{MN}dx^M dx^N=R^2\left(\frac{1}{4} ds^2_{AdS^4} +
ds^2_{\cp}\right),\h R^2=\sqrt{32\pi^2\lambda}, \eea where
$\lambda\equiv N/k$ is the 't Hooft coupling.
With $\alpha'=1$ convention, this coupling is related to the string tension by
\bea\nn
\frac{TR^2}{2}=\sqrt{2\lambda},
\eea
which is different from the case of $AdS_5\times S^5$.

The coordinates describing the background can be chosen such that
\bea\nn \sum_{i,j=0}^{4}\eta_{ij}y^i
y^j+\left(\frac{R}{2}\right)^2=0, \h
\eta_{ij}=diag(-1,1,1,1,-1),\eea for the $AdS$ part and \bea\nn
\sum_{i=1}^{8}(x^i)^2-R^2=0,\h \sum_{i=1,3,5,7}\left(x^i\p_m
x^{i+1}-x^{i+1}\p_m x^i\right)=0,\eea for the $\cp$ part \cite{Grignani}.
Further on, we restrict ourselves to the $R_t\times\cp$ subspace for which
$y^1=y^2=y^3=0$, and introduce the complex coordinates \bea\nn
z=y^0+iy^4,\h w_1=x^1+ix^2,\h w_2=x^3+ix^4,\h w_3=x^5+ix^6,\h
w_4=x^7+ix^8.\eea Now, we can embed the string as follows \bea\nn
&&z=Z(\tau,\sigma)=\frac{R}{2}e^{it(\tau,\sigma)},\h
w_a=W_a(\tau,\sigma)=R r_a(\tau,\sigma)e^{i\varphi_a(\tau,\sigma)}.
\eea
These complex coordinates should satisfy
\bea
\nn &&\sum_{a=1}^{4}W_a\bar{W}_a=R^2,\h
\eea
which corresponds to $S^7$ and further more
\bea
\sum_{a=1}^{4}\left(W_a\p_m\bar{W}_a-\bar{W}_a\p_m W_a\right)=0,
\label{embcp}
\eea
which reduces the embedding to $\cp$.
Here $t$ is the $AdS$ time.
In terms of the embedding coordinates, the $\cp$ condition (\ref{embcp}) becomes
\bea
\sum_{a=1}^{4}r_a^2\p_m\varphi_a=0,\quad m=0,1.
\label{cp3cond}
\eea
For this embedding, the metric induced
on the string worldsheet is given by \bea\nn
G_{mn}&=&-\p_{(m}Z\p_{n)}\bar{Z}
+\sum_{a=1}^{4}\p_{(m}W_a\p_{n)}\bar{W}_a
\\ \nn &=&R^2\left[-\frac{1}{4}\p_mt\p_nt +
\sum_{a=1}^{4}\left(\p_mr_a\p_nr_a +
r_a^2\p_m\varphi_a\p_n\varphi_a\right)\right].\eea The corresponding
string Lagrangian becomes \bea\nn \mathcal{L}=\mathcal{L}_s +
\sqrt{8\lambda}\Lambda\left(\sum_{a=1}^{4}r_a^2-1\right) +
\sqrt{8\lambda}\Lambda_0\sum_{a=1}^{4}r_a^2\p_0\varphi_a+\sqrt{8\lambda}\Lambda_1
\sum_{a=1}^{4}r_a^2\p_1\varphi_a,
\eea where $\Lambda$,
$\Lambda_0,\Lambda_1$ are Lagrange multipliers.

In the case at hand, the background metric does not depend on $t$
and $\varphi_a$. Therefore, the conserved quantities are the string
energy $E_s$ and four angular momenta $J_a$, given by
\bea\label{gcqs} E_s=-\int d\sigma\frac{\p\mathcal{L}}{\p(\p_0
t)},\h J_a=\int d\sigma\frac{\p\mathcal{L}}{\p(\p_0\varphi_a)}.\eea

In order to reduce the string dynamics on $R_t\times\cp$ to the NR integrable system,
we use the ansatz \cite{AFRT,ART,KRT06}
\bea\label{NRA} &&t(\tau,\sigma)=\kappa\tau,\h
r_a(\tau,\sigma)=r_a(\xi),\h
\varphi_a(\tau,\sigma)=\omega_a\tau+f_a(\xi),\\ \nn
&&\xi=\alpha\sigma+\beta\tau,\h \kappa, \omega_a, \alpha,
\beta={\rm constants}.\eea
Then the Lagrangian $\mathcal{L}$ takes the
form (prime is used for $\p/\p \xi$) \bea\nn
\mathcal{L}&=&-\sqrt{2\lambda}(\alpha^2-\beta^2)\sum_{a=1}^{4}
\left[r_a'^2+r_a^2\left(f_a'-\frac{\beta\omega_a}{\alpha^2-\beta^2}\right)^2
-\frac{\alpha^2\omega_a^2}{(\alpha^2-\beta^2)^2}r_a^2\right]
\\ \nn &+& \sqrt{8\lambda}\Lambda\left(\sum_{a=1}^{4}r_a^2-1\right) +
\sqrt{8\lambda}\Lambda_0\sum_{a=1}^{4}\omega_a r_a^2 + \sqrt{8\lambda}
\Lambda_1\sum_{a=1}^{4}f'_a r_a^2.\eea
Now we can integrate the equations of motion for $f_a$ to get
\bea\label{fafi} f'_a=\frac{1}{\alpha^2-\beta^2}
\left(\frac{C_a}{r_a^2}+\beta\omega_a+\Lambda_1\right), \h\eea
where $C_a$ are integration constants. By using (\ref{fafi}), the
equations of motion for $r_a$ can be written as \bea\nn
\left(\alpha^2-\beta^2\right)r^{''}_a - \frac{1}{\alpha^2-\beta^2}
\frac{C_a}{r_a^3}
+\left[\omega_a^2+2\left(\Lambda+\Lambda_0\omega_a\right)
+\frac{\left(\Lambda_1+\beta\omega_a\right)^2}{\alpha^2-\beta^2}
\right]r_a=0.\eea These can be
obtained from the Lagrangian \bea\nn
L&=&\sum_{a=1}^{4}\left[(\alpha^2-\beta^2)r^{'2}_a
-\frac{1}{\alpha^2-\beta^2} \frac{C_a}{r_a^2}-\omega_a^2r_a^2\right]
\\ \nn &-&2\Lambda\left(\sum_{a=1}^{4}r_a^2-1\right) -2
\Lambda_0\sum_{a=1}^{4}\omega_a r_a^2
-\frac{1}{\alpha^2-\beta^2}\sum_{a=1}^{4}
\left(\Lambda_1+\beta\omega_a\right)^2r_a^2.\eea From the
equations of motion for the Lagrange multipliers it follows that
$\Lambda_1=0$. Thus, we end up with the following effective
Lagrangian for the coordinates $r_a$ \bea \label{NRL}
L_{NR}&=&(\alpha^2-\beta^2)
\sum_{a=1}^{4}\left[r^{'2}_a-\frac{1}{(\alpha^2-\beta^2)^2}
\left(\frac{C_a^2}{r_a^2} + \alpha^2\omega_a^2r_a^2\right)\right]
\\ \nn &-&2\Lambda\left(\sum_{a=1}^{4}r_a^2-1\right) -2
\Lambda_0\sum_{a=1}^{4}\omega_a r_a^2.\eea
This is the Lagrangian for the NR integrable system \cite{KRT06}
with one more embedding condition which comes from the $\cp$ condition (\ref{cp3cond}),
\bea
\sum_{a=1}^{4}\omega_a r_a^2=0.\label{aec}
\eea
In addition, from (\ref{fafi}) and the constraint
$\sum_{a=1}^{4}f'_a r_a^2=0$, one can find
$\Lambda_1=-\sum_{a=1}^{4}C_a$.
Since $\Lambda_1$ should be zero, this leads to
\bea\label{Cac}
\sum_{a=1}^{4}C_a=0.
\eea
These two extra conditions for the NR system of $\cp$ are the main difference from
that of the sphere geometry.
In other words, strings moving on $R_t\times \cp$ should satisfy these two
conditions additionally.

The Virasoro constraints (\ref{00}) give the conserved Hamiltonian $H_{NR}$
and a relation between the embedding parameters and the arbitrary
constants $C_a$: \bea\label{HNR} &&H_{NR}=(\alpha^2-\beta^2)
\sum_{a=1}^{4}\left[r_a'^2+\frac{1}{(\alpha^2-\beta^2)^2}
\left(\frac{C_a^2}{r_a^2} + \alpha^2\omega_a^2r_a^2\right)\right]
=\frac{\alpha^2+\beta^2}{\alpha^2-\beta^2}\frac{\kappa^2}{4},
\\ \label{01R} &&\sum_{a=1}^{4}C_a\omega_a + \beta(\kappa/2)^2=0.\eea
For closed strings, $r_a$ and $f_a$ satisfy the following
periodicity conditions \bea r_a(\xi+2\pi\alpha)=r_a(\xi),\h
f_a(\xi+2\pi\alpha)=f_a(\xi)+2\pi n_a,\label{pbc} \eea where $n_a$
are integer winding numbers.

The conserved charges can be computed from the definition (\ref{gcqs}).
Using the ansatz (\ref{NRA}), one can express the angular momenta as
\bea
J_a=\frac{2\sqrt{2\lambda}}{\alpha}\int d\sigma\ r_a(\xi)^2\p_0\varphi_a,\quad a=1,2,3,4.
\eea
Inserting the solutions (\ref{fafi}) into these, we can find
\bea\label{cqs} E_s=
\frac{\kappa\sqrt{2\lambda}}{2\alpha}\int d\xi,\h J_a=
\frac{2\sqrt{2\lambda}}{\alpha^2-\beta^2}\int d\xi
\left(\frac{\beta}{\alpha}C_a+\alpha\omega_a r_a^2\right).\eea In
view of (\ref{aec}) and (\ref{Cac}), one arrives at \bea\label{Jc}
\sum_{a=1}^{4}J_a=0.\eea
This condition, first noticed in \cite{Grignani}, appears naturally in our
NR approach.

\setcounter{equation}{0}
\section{Two Angular Momenta Solutions}

In this section we are interested in finding string configurations
corresponding to the following particular solution of (\ref{aec}) and (\ref{Cac})
\bea\nn r_1=r_3,\h r_2=r_4,\h \omega_1=-\omega_3,\h
\omega_2=-\omega_4.\eea
Two angular velocities $\omega_1,\omega_2$ are independent and
lead to strings moving in $\cp$ with two angular momenta.
A special case $\omega_2=0$ corresponds to the cases considered in \cite{Grignani,Grignanif}.

\subsection{Explicit Solutions}

We will use the parametrization \bea\nn
r_1=r_3=\frac{1}{\sqrt{2}}\sin\theta,\h
r_2=r_4=\frac{1}{\sqrt{2}}\cos\theta.\eea From the NR Hamiltonian
(\ref{HNR}) one finds
\bea\nn
\theta'^2(\xi)=\frac{1}{(\alpha^2-\beta^2)^2}
\left[\frac{\kappa^2}{4}(\alpha^2+\beta^2) -
2\left(\frac{C_1^2+C_3^2}{\sin^2{\theta}} +
\frac{C_2^2+C_4^2}{\cos^2{\theta}}\right)-
\alpha^2\left(\omega_1^2\sin^2{\theta}
+\omega_2^2\cos^2{\theta}\right)\right]. \eea
We further restrict ourselves to $C_2=C_4=0$ to search for GM and SS solutions.
Eqs.(\ref{Cac}) and (\ref{01R}) give
\bea\nn
C_1=-C_3=-\frac{\beta\kappa^2}{8\omega_1}.
\eea

In this case, the above
equation for $\theta'$ can be rewritten in the form
\bea\label{tS3eq}
(\cos\theta)'=\mp\frac{\alpha\sqrt{\omega_1^2-\omega_2^2}}{\alpha^2-\beta^2}
\sqrt{(z_+^2-\cos^2\theta)(\cos^2\theta-z_-^2)},\eea where \bea\nn
&&z^2_\pm=\frac{1}{2(1-\frac{\omega_2^2}{\omega_1^2})}
\left\{y_1+y_2-\frac{\omega_2^2}{\omega_1^2}
\pm\sqrt{(y_1-y_2)^2-\left[2\left(y_1+y_2-2y_1
y_2\right)-\frac{\omega_2^2}{\omega_1^2}\right]
\frac{\omega_2^2}{\omega_1^2}}\right\}, \\ \nn
&&y_1=1-\frac{\kappa^2}{4\omega_1^2},\h
y_2=1-\frac{\beta^2}{\alpha^2}\frac{\kappa^2}{4\omega_1^2}.\eea
The solution of
(\ref{tS3eq}) is given by \bea\label{tS3sol} \cos\theta=z_+
dn\left(C\xi|m\right),\h
C=\mp\frac{\alpha\sqrt{\omega_1^2-\omega_2^2}}{\alpha^2-\beta^2}
z_+,\h m\equiv 1-z^2_-/z^2_+ ,\eea where $dn\left(C\xi|m\right)$ is
one of the elliptic functions.

To find the full string solution, we also need to obtain the
explicit expressions for the functions $f_a$ from (\ref{fafi})
\bea\nn
f_a=\frac{1}{\alpha^2-\beta^2}\int
d\xi\left(\frac{C_a}{r_a^2}+\beta\omega_a\right).\eea
Using the solution for $\theta(\xi)$ in (\ref{tS3sol}), we can find
\bea\nn
&&f_1=-f_3=\frac{\beta/\alpha}{z_+\sqrt{1-\omega_2^2/\omega_1^2}}
\left[C\xi -
\frac{2(\kappa/2)^2/\omega_1^2}{1-z^2_+}\Pi\left(am(C\xi),-\frac{z^2_+
-z^2_-}{1-z^2_+}|m\right)\right],\\ \nn
&&f_2=-f_4=\frac{\beta\omega_2}{\alpha^2-\beta^2}\xi.\eea
Here, $\Pi$ is the elliptic integrals of the third kind. As a
consequence, the full string solution is given by \bea \nn
&&W_1=\frac{R}{\sqrt{2}}\sqrt{1-z_+^2dn^2\left(C\xi|m\right)}\
e^{i(\omega_1\tau+f_1)},\\ \label{fss} &&W_2=\frac{R}{\sqrt{2}}z_+
dn\left(C\xi|m\right)\ e^{i(\omega_2\tau+f_2)},\\ \nn
&&W_3=\frac{R}{\sqrt{2}}\sqrt{1-z_+^2dn^2\left(C\xi|m\right)}\
e^{-i(\omega_1\tau+f_1)},\\ \nn &&W_4=\frac{R}{\sqrt{2}}z_+
dn\left(C\xi|m\right)\ e^{-i(\omega_2\tau+f_2)} .\eea  Let us also
note that (\ref{fss}) contains both cases: $\alpha^2>\beta^2$ and
$\alpha^2<\beta^2$, which correspond to the GM and SS strings
respectively.

The geometric meaning of the explicit solutions (\ref{fss}) is as follows.
Each pairs of complex coordinates, $(W_1,W_2)$ and $(W_3,W_4)$, describe
a spiky solutions in $S^2$ sphere geometry but with dynamics at opposite points in the $U(1)$ fiber.
The two phases in $(W_1,W_2)$ are exactly opposite to those of $(W_3,W_4)$ which, together with
the dynamivs in $U(1)$, ensures vanishing of the total momentum.
This behavior has been also noticed for the string in $R_t\times S^2\times S^2$ in
\cite{Grignani}.

\subsection{Infinite volume limit}
The GM and SS in the infinite volume can be obtained by taking $z_-\to 0$.
In this limit, the solution reduces to
\bea\nn
\cos\theta=\frac{\sin\frac{p}{2}}{\cosh(C\xi)},
\eea
where the constant $z_{+}\equiv\sin p/2$ is given by
\bea\nn
z^2_{+}=\frac{y_2-\omega_2^2/\omega_1^2}{1-\omega_2^2/\omega_1^2}\qquad({\rm GM}),
\qquad{\rm and}\qquad z^2_{+}=\frac{y_1-\omega_2^2/\omega_1^2}{1-\omega_2^2/\omega_1^2}
\qquad({\rm SS}).
\eea
One angular momentum solutions are given by $\omega_2=0$.
Inserting these into the energy and angular momenta expressions in (\ref{cqs}),
we can find the following energy-charge dispersion relation:
\bea
E_s-J_1=\sqrt{J_2^2+8\lambda\sin^2\frac{p}{2}}.
\label{infdis}
\eea
While this is exactly same as that of the dyonic GM
in the $AdS_5\times S^5$, the result arises from quite different string dynamics in the $\cp$.

\subsection{Finite-size effects}
Using the most general solutions (\ref{fss}), we can calculate
the finite-size corrections to the energy-charge relation (\ref{infdis}) in
the limit when the string energy $E_s\to\infty$.
This analysis depends crucially on the sign of the
difference $\alpha^2-\beta^2$.
The GM solution corresponds to $\alpha^2>\beta^2$ while the SS
to $\alpha^2<\beta^2$.
While the string dynamics are quite different, computations are identical to
the cases in the sphere geometries.
Therefore, we will provide only the results here, referring
technical details to \cite{ABi,ABii}.

\subsubsection{Giant magnon}

We begin with the GM case, i.e. $\alpha^2>\beta^2$. Then, one
obtains from (\ref{cqs}) the following expressions for the conserved
string energy $E_s$ and the angular momenta $J_a$ \bea\nn
&&\mathcal{E} =\frac{2\kappa(1-\beta^2/\alpha^2)} {\omega_1
z_+\sqrt{1-\omega_2^2/\omega_1^2}}\mathbf{K}
\left(1-z^2_-/z^2_+\right), \\ \label{cqsGM} &&\mathcal{J}_1=
\frac{2 z_+}{\sqrt{1-\omega_2^2/\omega_1^2}} \left[
\frac{1-\beta^2(\kappa/2)^2/\alpha^2\omega_1^2}{z^2_+}\mathbf{K}
\left(1-z^2_-/z^2_+\right)-\mathbf{E}
\left(1-z^2_-/z^2_+\right)\right], \\ \nn &&\mathcal{J}_2= \frac{2
z_+ \omega_2/\omega_1 }{\sqrt{1-\omega_2^2/\omega_1^2}}\mathbf{E}
\left(1-z^2_-/z^2_+\right),\h \mathcal{J}_3=-\mathcal{J}_1,\h
\mathcal{J}_4=-\mathcal{J}_2.\eea As a result, the condition
(\ref{Jc}) is identically satisfied. Here, we introduced the
notations \bea\label{not} \mathcal{E}=\frac{E_s}{\sqrt{2\lambda}}
,\h \mathcal{J}_a=\frac{J_a}{\sqrt{2\lambda}}.\eea The computation
of $\Delta\varphi_1$ gives \bea\label{pws} p\equiv\Delta\varphi_1
&=& 2\int_{\theta_{min}}^{\theta_{max}}\frac{d \theta}{\theta'}f'_1=
\\ \nn &-&\frac{2\beta/\alpha}{z_+\sqrt{1-\omega_2^2/\omega_1^2}}
\left[\frac{(\kappa/2)^2/\omega_1^2}{1-z^2_+}\Pi\left(-\frac{z^2_+ -
z^2_-}{1-z^2_+}\bigg\vert 1-z^2_-/z^2_+\right) -\mathbf{K}
\left(1-z^2_-/z^2_+\right)\right].\eea In the above expressions,
$\mathbf{K}(m)$, $\mathbf{E}(m)$ and $\Pi(n|m)$ are the complete
elliptic integrals.

Expanding the elliptic integrals, we obtain
\bea\label{IEJ1} &&\mathcal{E}-\mathcal{J}_1 =
\sqrt{\mathcal{J}_2^2+4\sin^2(p/2)} - \frac{16 \sin^4(p/2)}
{\sqrt{\mathcal{J}_2^2+4\sin^2(p/2)}}\times\\ \nn
&&\exp\left[-\frac{2\left(\mathcal{J}_1 +
\sqrt{\mathcal{J}_2^2+4\sin^2(p/2)}\right)
\sqrt{\mathcal{J}_2^2+4\sin^2(p/2)}\sin^2(p/2)}{\mathcal{J}_2^2+4\sin^4(p/2)}
\right].\eea
It is easy to check that the energy-charge relation
(\ref{IEJ1}) coincides with the results in \cite{HS08} (and in \cite{AFZ} for $J_2=0$),
which describes the finite-size effects for dyonic GM on $R_t\times S^3$ subspace of
$AdS_5\times S^5$. The difference is that in the last case the
relations between $\mathcal{E}$, $\mathcal{J}_1$, $\mathcal{J}_2$
and $E$, $J_1$, $J_2$ are given by \bea\nn
\mathcal{E}=\frac{2\pi}{\sqrt{\lambda}}E ,\h
\mathcal{J}_1=\frac{2\pi}{\sqrt{\lambda}}J_1, \h
\mathcal{J}_2=\frac{2\pi}{\sqrt{\lambda}}J_2,\eea while for strings
on $R_t\times\cp$ they are written in (\ref{not}).

\subsubsection{Single spike}

Now, we turn our attention to the SS case, when $\alpha^2<\beta^2$.
The computation of the conserved quantities (\ref{cqs}) and
$\Delta\varphi_1$ gives \bea\nn &&\mathcal{E}
=\frac{\kappa(\beta^2/\alpha^2-1)}
{\omega_1\sqrt{1-\omega_2^2/\omega_1^2}z_+}\mathbf{K}
\left(1-z^2_-/z^2_+\right), \\ \nn &&\mathcal{J}_1= \frac{2
z_+}{\sqrt{1-\omega_2^2/\omega_1^2}} \left[\mathbf{E}
\left(1-z^2_-/z^2_+\right)
-\frac{1-\beta^2(\kappa/2)^2/\alpha^2\omega_1^2}{z^2_+}\mathbf{K}
\left(1-z^2_-/z^2_+\right)\right], \\ \nn &&\mathcal{J}_2= -\frac{2
z_+ \omega_2/\omega_1 }{\sqrt{1-\omega_2^2/\omega_1^2}}\mathbf{E}
\left(1-z^2_-/z^2_+\right), \\ \nn &&\Delta\varphi_1=
-\frac{2\beta/\alpha}{\sqrt{1-\omega_2^2/\omega_1^2}z_+}
\left[\frac{(\kappa/2)^2/\omega_1^2}{1-z^2_+}\Pi\left(-\frac{z^2_+ -
z^2_-}{1-z^2_+}|1-z^2_-/z^2_+\right) -\mathbf{K}
\left(1-z^2_-/z^2_+\right)\right].\eea
From these, we obtain
\bea\nn \mathcal{J}_1=\sqrt{\mathcal{J}_2^2+4\sin^2(p/2)},\eea
and
\bea\label{ssS3c} \mathcal{E}-\Delta\varphi_1=
p+8\sin^2\frac{p}{2}\tan\frac{p}{2}
\exp\left(-\frac{\tan\frac{p}{2}(\Delta\varphi_1 + p)}
{\tan^2\frac{p}{2} + \mathcal{J}_2^2 \csc^2p}\right).\eea
This is the leading finite-size correction to the
``$E-\Delta\varphi$'' relation for the SS string with two angular
momenta on $R_t\times\cp$. It coincides with the string result for
$R_t\times S^3$ found in \cite{ABi}. As in the GM case, the
difference is in the identification (\ref{not}).

\section{Concluding Remarks}
We have shown that the NR integrable system is particularly effective
to find classical string solutions for $AdS_4\times\cp$.
The extra constraints arising from $\cp$ geometry can be naturally reformulated into
simple conditions under the NR ansatz.
In addition to the GM and SS solutions moving in $R_t\times S^2\times S^2$
with a single angular momentum,
the NR system can be used to study more complicated string dynamics.
As shown in this paper, the GM and SS solutions in $\cp$ with two angular momenta solutions
can be described in the same way as the cases in the sphere geometries.
These solutions describe the strings moving in $R_t\times\cp$ where
the two angular momenta in one $S^2$ are opposite to those in the other $S^2$ executing
motion on $S^1$ in the $U(1)$ Hopf fibration over $S^2\times S^2$.
Of course, the extra constraints are limiting the possible string configurations.
It would be interesting to find other configurations which could be found within
the context of the NR system (some solutions are given in the Appendix).
Another interesting feature would be the relation of the NR integrable system
to other classical integrable systems such as complex sine-Gordon model
as has been shown for the type IIB string theory on $AdS_5\times S^5$ in \cite{ABi}.

The two angular momenta string states are related to the composite
operators in the gauge theory side.
The BPS state corresponding to the string vacuum is
${\rm tr}\left[ (A_{1} B_{1})^L\right]$
where $A_{1}$ and $B_{1}$ are the scalar fields of ${\cal N}=6$ Chern-Simons
theory in the bifundamental representation
$({\bf N}, {\bf \bar N})$ and $({\bf \bar N}, {\bf N})$, respectively.
The excited states are obtained by replacing these fields with fields in the theory.
The composite operators dual to the string with two angular momenta with the
dispersion relation (\ref{infdis}) should be
\bea\nn
{\rm tr}\left[ (A_{1} B_{1})^{J_1}(A_{2} B_{2})^{J_2}\right]+\ldots \,,
\eea
where the ellipsis represents the permutations of the fields while maintaining the
alternating spin chain structure and
$A_2$ and $B_2$ are another scalar fields.

It would be interesting to compare the energy-charge dispersion relation
we have obtained here with the solutions of all-loop Bethe ansatz equations
recently proposed in \cite{GV}.
The finite-size corrections can not be derived solely from the Bethe ansatz equations.
Instead, one can use the L\"uscher correction formulation based on exact $S$-matrix
which has been particularly powerful in the AdS/CFT correspondence \cite{BajJan}.
It would be interesting to compute the corrections based on a recently proposed
$S$-matrix \cite{AN} and compare with our results in the large `t Hooft coupling limit.

\section*{Acknowledgements}
This work was supported in part by KRF-2007-313-C00150 (CA), by NSFB
VU-F-201/06 (PB, RR), and by the Brain Pool program 2007-1822-1-1 from the Korean
Federation of Science and Technology (PB).

\section*{Appendix: More String Solutions on $R_t\times\cp$}
\def\theequation{A.\arabic{equation}}
\setcounter{equation}{0}
\begin{appendix}

\paragraph{Basic facts about $\cp$:}

Let us explain briefly the basic properties of the $\cp$ spaces.
It is most convenient to define an $n$-dimensional complex projective space $\mathbb {CP}^n$
as the family of one-dimensional subspaces
in $\mathbb{C}^{n+1}$, i.e. this is the quotient $\mathbb{C}^{n+1}/(\mathbb{C}\setminus\{0\})$.
The equivalence relation is defined as
\bea\nn
\a Z_1:\cdots:\a Z_{n+1}=Z_1:\cdots:Z_{n+1}.\notag
\eea
The space $\mathbb {CP}^n$ itself is covered by patches
$U_i:\{Z_1:\cdots:Z_{n+1}\in\mathbb{CP}^n\,|\,Z_i\neq 0\}, \,\,
i=1,\cdots,n+1$. One can see that each patch $U_i$ is isomorphic to $\mathbb {CP}^n$,
where the isomorphism is defined
by $W_j^{(i)}=Z_j/Z_i,\,\, j\neq i$. One can choose local coordnates
$W=(W_1,W_2,\cdots,W_n)^t\in\mathbb{C}^{n+1}$ with
$W_j\equiv W_j^{(n+1)}$. The Fubini-Study metric then is given by the line element
\bea\nn
ds^2=\frac{(1+|W|^2)|dW|^2-|W^\dagger dW|^2}
{(1+|W|^2)^2}.
\eea

One can think of $\mathbb {CP}^n$ as the homogeneous space
$\mathbb {CP}^n=U(n+1)/(U(n)\times U(1))$. The
$u(n+1)$ Lie algebra $\frak{f}$ can be realized as anti-hermitian matrices
and splits into two parts: $\frak{p}=u(n)\oplus u(1)$
and its orthogonal completion $\frak{cp}(n)$ with respect to the $U(n+1)$ Killing for
\bea\nn
& &\frak{p}=u(n)\oplus u(1)=\{iM\in u(n+1)\,|\,[\Gamma,M]=0\} \notag \\
& &\frak{cp}(n)=\{iM\in u(n+1)\,|\,\{\Gamma,M\}=0\},\notag
\eea
where $M$ is traceless and hermitian and
\bea\nn
\Gamma=\begin{pmatrix}\mathbf{1}_n & \\ & -1 \end{pmatrix}.
\eea
A generator of $\frak{cp}(n)$ part, $\mathbf{B}$ then is given by
\bea\nn
\mathbf{B}=\begin{pmatrix}
& W^\dagger \\ -W \end{pmatrix}.
\eea
Then one can write schematically
\bea\nn
 \frak{f}=\frak{p}\oplus \frak{cp}, \quad [\frak{p},\frak{p}]\subset \frak{p}, \quad
[\frak{p},\frak{cp}]\subset \frak{cp}, \quad [\frak{cp},\frak{cp}]\subset \frak{cp}.
\eea

\paragraph{More strings on $R_t\times\cp$ and NR system:}

The next step is to consider concrete solutions, taking into account
all of the existing constraints, which can be summarized as follows
\bea\nn \sum_{a=1}^{4}r_a^2=1,\h  \sum_{a=1}^{4}\omega_a r_a^2=0, \h
\sum_{a=1}^{4}C_a=0,\h \sum_{a=1}^{4}C_a\omega_a +
\beta(\kappa/2)^2=0.\eea From the first two equalities we can
express two of the $r_a$ coordinates through the remaining ones.
Then, we are left only with relations between the parameters. In
order to be able to compare with the known particular solutions, we
choose to express $r_{1,3}^2$ as functions of $r_{2,4}^2$. The
general solution is \bea\label{gsfr}
r_1^2=\frac{\omega_3(1-r_2^2-r_4^2) +\omega_2 r_2^2+\omega_4
r_4^2}{\omega_3-\omega_1},\h r_3^2=\frac{\omega_1(1-r_2^2-r_4^2)
+\omega_2 r_2^2+\omega_4 r_4^2}{\omega_1-\omega_3}.\eea In
particular, for $\omega_1=-\omega_3$, $\omega_2=\omega_4=0$, we have
$r_1^2=r_3^2$. The case considered in \cite{Grignani,Grignanif} is
reached after fixing $r_2^2=r_4^2$, $r_1^2+r_2^2=1/2$.

Denoting dynamical variables as
\bea\nn
r_1=r_3=\frac{r}{\sqrt{2}}, \quad r_2=r_4=\frac{\sqrt{1-r^2}}{\sqrt{2}},
\eea
one can describe the system  by only one independent variable $r$.
The first order differential equation for $r$ can be obtained
either from the equations of motion (integrating them
once) or from the Virasoro constraints.
It goes as follows
\bea
& \sum\limits_a\lb[(\a^2-\b^2){r_a'}^2+\frac{C_a^2}{\a^2-\b^2}\frac{1}{r_a^2}
+\frac{\a^2}{\a^2-\b^2}\omega_a^2r_a^2+\frac{2\b
C_a\omega_a}{\a^2-\b^2}\rb]=(\kappa/2)^2 \notag \\
& \Rightarrow\quad
(1-\b^2)\frac{{r'}^2}{1-r^2}+\frac{4}{1-\b^2}\frac{C^2}{r^2}+
\frac{1}{1-\b^2}\o^2 r^2+\frac{4\b C\o}{1-\b^2}=(\kappa/2)^2
\label{mag-1}
\eea
Here, without loss of generality, we set $\a=1$. Using the constraint
\bea\nn
2C\o+\b(\kappa/2)^2=0 \notag
\eea
we find
\bea
(1-\b^2)^2{r'}^2
& =(1-r^2)\lb\{(1+\b^2)(\kappa/2)^2-\frac{4C^2}{r^2}-\o^2r^2\rb\} \notag \\
& =-\omega^2\frac{(1-r^2)}{r^2}\lb\{\frac{4C^2}{\o^2}- (1+\b^2)
\frac{(\kappa/2)^2}{\o^2}r^2+r^4\rb\}
\label{eq-r-2sp}
\eea
The right hand side determines the turning
points ${r'}^2=0$ and they are three. In order the string to
extends to the equator of the sphere, one must choose $r=1$.
 To find a
solution of the type we are looking for, $r^2=1$ has to be double
zero of the right hand side of (\ref{eq-r-2sp}). The latter
conditions leads to the following constraints
\bea\nn
(1+\b^2)(\kappa/2)^2=\omega^2+4C^2, \quad 2 C\omega
+\b(\kappa/2)^2=0
\label{eq-param-nr}
\eea
which can be obtained either
by substituting $r=1$ in the right hand side of (\ref{eq-r-2sp})
or from the Virasoro constraints. The correct choice for the
parameters solving the above equation and giving GM type
string solutions is
\bea
\kappa/2=\omega, \quad \a=1, \quad
\b=-\frac{2C}{\omega}
\label{param-2sp}
\eea

Let us turn to the solutions developing a SS in the $R_t\times S^2\times S^2$
subspace.
This configuration can be realized in terms of the NR integrable
system with specific choice of the parameters. The solutions we
are looking for are characterized by large quantum numbers,
especially large energy. The careful analysis shows that in order to have
such solutions one has to choose the parameters in a specific way.
The ``spiky'' choice for the parameters, namely the choice giving
solutions with a SS but infinitely wound around the
equator, is slightly different from the case of the GM. In
fact the constraints on the parameters are the same,
but instead of the choice (\ref{param-2sp}),
now we choose the other solution to the constraint
\bea\nn
\kappa=2C, \quad
\beta=-\frac{2\omega C}{\kappa^2}=-\frac{\omega}{2C}
\label{param-ssp}
\eea
The equation for the variable $r$ is the same, but the parameters are
fixed differently
\bea\nn
\frac{d\,u}{d\xi}=u'=\frac{2\omega}{1-\beta^2}(1-u)\sqrt{u-\bar u}.
\eea
Above we use the following notations
\bea\nn
u&=&r^2=\sin^2\theta, \quad \bar
u=\frac{4C^2}{\omega^2},   \notag \\
d\xi&=&\frac{du}{u'}=\frac{(1-\beta^2)\,du}{2\omega(1-u)
\sqrt{u-\bar u}}= \frac{(4C^2-\omega^2)\,d
u}{8C^2\omega(1-u)\sqrt{u-\bar u}}
\eea
The conserved quantities are
\bea\nn
E&=&\kappa T\int d\xi \notag \\
J&=&\frac{C\beta}{(1-\beta^2)}\int d\xi + \frac{\omega}{(1-\beta^2)}T\int u\,d\xi \notag\\
\Delta\phi&=&\frac{C}{(1-\beta^2)}\int\frac{d\xi}{u} +
\frac{\beta\omega}{(1-\beta^2)}\int d\xi. \notag
\eea

To find finite results (which is so for $E-J$ in the GM case) we
consider
\bea\nn
E-T\Delta\phi=\frac{2CT}{\omega}\arccos\sqrt{\bar
u}=\frac{\sqrt{\lambda}}{\pi}\bar\theta
\eea
where
\bea\nn
\bar\theta=\frac{\pi}{2}-\theta_0
\eea
For the total spin $J$ we get
\bea\nn
J=\frac{2T\omega}{\omega}\cos\theta_0=2T\sin\bar\theta
\eea
All this implies finally
\bea\nn
\Delta=(E-T\Delta\phi)-J=\frac{\sqrt{\lambda}}{\pi}(\bar\theta-\sin\bar\theta)
\eea
which completes our result on SS case with this ansatz.
We note that the solution in this case is
\bea
\sin\theta=\tanh\lb(\frac{\o\bar z}{1-\b^2}(\xi-\xi_0)\rb).
\label{sol-2}
\eea

Now we consider more general solutions.
It is reasonable to ask for symmetric motion on the subspace $S^2\times S^2$ and therefore
to set
\bea\nn
r_1^2+r_2^2=\frac{1}{2}, \quad r_3^2+r_4^2=\frac{1}{2}.
\eea
Then one can use the parametrization
\bea\nn
r_1^2+r_3^2=r^2, \quad r_2^2+r_4^2=1-r^2.
\eea
Having in mind that the total worldshhet momentum has to be zero, one can set
\bea\nn
C_1=-C_3, \quad \o_1=-\o_3=\o.
\eea
It follows then that
\bea\nn
C_2=-C_4.
\eea
Using the above constraints one can find
\bea\nn
\lb(1-\frac{\o_2}{\o}\rb)r_2^2-\lb(\frac{\o_4}{\o}+1\rb)r_4^2=0.
\eea
There are two cases:
a) $\o_2=\o$ which entails $\o_4=-\o$ -- this choice slightly generalizes the case
considered above;
b) $r_2$ and $r_4$ are proportional
\bea\nn
r_2^2=\frac{\o_4+\o}{\o-\o_2}r_4^2:=\Gamma^2r_4^2.
\eea
Let us consider the last possibility in more details.
The constraints tells us that
\bea\nn
r_2^2&=&\frac{\Gamma^2}{1+\Gamma^2}(1-r^2), \quad
r_4^2=\frac{1}{1+\Gamma^2}(1-r^2) \notag \\
r_1^2&=&\frac{\o(1+\Gamma^2)+\o_2\Gamma^2+\o_4}{\o(1+\Gamma^2)}r^2
-\frac{\o_2\Gamma^2+\o_4}{\o(1+\Gamma^2)}.
\eea
Rewriting the last equality as
\bea\nn
r_1^2=(1+b^2)r^2-b^2,\quad b^2=\frac{\o_2\Gamma^2+\o_4}{\o(1+\Gamma^2)},
\eea
one can find the lower bound
\bea\nn
r_{min}\leq r, \quad r_{min}=\frac{b}{\sqrt{1+b^2}}.
\eea

It is better to use $y^2=(1+b^2)x^2=(1+b^2)(1-r^2)$ and then the radial dynamical
variables become
\bea
r_1^2=1-y^2,\quad r_2^2=c^2y^2,\quad r_3^2=d^2y^2, \quad r_4^2=(1-c^2)y^2,\label{rename}
\eea
where
\bea\nn
c^2=\frac{\Gamma^2}{(1+b^2)(1+\Gamma^2)}, \quad
d^2=\frac{b^2}{(1+b^2)}. \notag
\eea
The Virasoro constraints give
\bea
(\a^2-\b^2)\frac{y'{}^2}{1-y^2}+\frac{1}{\a^2-\b^2}\lb[\frac{C_1^2}{1-y^2}
+\frac{\tilde C^2}{y^2}\rb]
+\frac{\a^2}{\a^2-\b^2}\lb[\o_1^2+\tilde\o^2y^2\rb]=\frac{\a^2+\b^2}{\a^2-\b^2}(\kappa/2)^2,
\label{eq-p1}
\eea
where
\bea\nn
\tilde C^2=\tilde C_2^2+\tilde C_3^2+\tilde C_4^2, \quad
\tilde C_2^2=\frac{C_2^2(1+b^2)}{c^2}, \quad
\tilde C_3^2=\frac{C_3^2(1+b^2)}{b^2}, \quad
\tilde C_4^2=\frac{C_4^2(1+b^2)}{1-c^2}.\notag
\eea
The system we obtained has the same type solutions as in the previous considerations and
can be solved in terms of elliptic functions.
Note that all the spins are different, so we find multi-spin solutions.

It is a standard procedure to bring the above equation into a Weierstrass form.
To do that we define
\bea\nn
\tilde\xi=\frac{\tilde\omega}{1-\b^2}\xi,\quad \zeta=\frac{a}{3}+y^2,
\eea
and rewrite the equation \eqref{eq-p1} as
\eq{
\frac{d\zeta}{d\tilde\xi}=4\zeta^3-g_2\zeta-g_3,
\label{eq-multi1}
}
with
\bea\nn
g_2=\frac{\tilde a^2}{3}-\tilde b ,\quad g_3=\tilde C^2=\frac{\tilde a\tilde b}{3}
+\frac{2}{27}\tilde a^3,\eea
where
\bea\nn
\tilde a=\frac{\omega_1^2}{\tilde\omega^2}-1-\frac{1+\b^2}{\tilde\omega^2}(\kappa/2)^2,
\quad \tilde b=\frac{1+\b^2}{\tilde\omega^2}(\kappa/2)^2
+\frac{\tilde C^2-\omega_1^2-C_1^2}{\tilde\omega^2}.
\eea
The solution is
\eq{
\zeta=e_3-e_{31}\mathrm{dn}^2(\sqrt{e_{31}\tilde\xi},\tilde\kappa)
\label{sol-multi1}
}
where $e_i$ are the roots of the rhs of \eqref{eq-multi1}, $e_{mn}=e_m-e_n$ and the modulus
is defined by ${\tilde\kappa}=e_{21}/e_{31}$. Going back to the variable $r$ we get
\bea\nn
r^2=1-\frac{1}{1+\b^2}(e_3-\frac{a}{3})+
\frac{e_{31}}{1+\b^2}\mathrm{dn}^2(\sqrt{e_{31}}\frac{\tilde\omega}{1-\b^2}\xi,\kappa).
\eea
Making specific choice of the parameters as in the above, one can get either GM
or SS solutions. The dispersion relations can be obtained using the explicit
form of the charges \eqref{cqs} and the relevant constraints.
\end{appendix}

\end{document}